\documentclass[showpacs,preprintnumbers,amsmath,amssymb]{revtex4}

\usepackage{graphicx}
\usepackage{dcolumn}
\usepackage{bm}
\usepackage{url}
\usepackage{epsfig}
\usepackage{multirow}
\usepackage{tensor}

\def\be{\begin{equation}}
\def\ee{\end{equation}}
\def\ba{\begin{eqnarray}}
\def\ea{\end{eqnarray}}

\newcommand{\fr}[2]{\frac{#1}{#2}}

\def\m{\rm{m}}

\def\Omo{\Omega_{\rm{m}0}}

\newcommand{\JCAP}{J.\ Cosmol.\ Astropart.\ Phys.}
\newcommand{\MNRAS}{Mon.\ Not.\ Roy.\ Astron.\ Soc.}

\def\ga{\mathrel{\raise.3ex\hbox{$>$\kern-.75em\lower1ex\hbox{$\sim$}}}}
\def\la{\mathrel{\raise.3ex\hbox{$<$\kern-.75em\lower1ex\hbox{$\sim$}}}}

\begin{document}

\title{Measuring the matter energy density and Hubble parameter from Large Scale Structure}


\author{Seokcheon Lee}
\affiliation{School of Physics, Korea Institute for Advanced Study, Heogiro 85, Seoul 130-722, Korea}


\begin{abstract}
We investigate the method to measure both the present value of the matter energy density contrast and the Hubble parameter directly from the measurement of the linear growth rate which is obtained from the large scale structure of the Universe. From this method, one can obtain the value of the nuisance cosmological parameter $\Omo$ (the present value of the matter energy density contrast) within $3$ \% error if the growth rate measurement can be reached $z > 3.5$. One can also investigate the evolution of the Hubble parameter without any prior on the value of $H_0$ (the current value of the Hubble parameter). Especially, estimating the Hubble parameter are insensitive to the errors on the measurement of the normalized growth rate $f \sigma_8$. However, this method requires the high $z$ ($z > 3.5$) measurement of the growth rate in order to get the less than $5$ \% errors on the measurements of $H(z)$ at $z \leq 1.2$ with the redshift bin $\Delta z = 0.2$. Thus, this will be suitable for the next generation large scale structure galaxy surveys like WFMOS and LSST.

\end{abstract}


\maketitle

\section{Introduction}
\setcounter{equation}{0}
The determination of the Hubble parameter, $H(z)$ has a practical and theoretical importance to many astrophysical properties and cosmological observations like cosmological rulers, the big bang nucleosynthesis (BBN), and the age of the Universe \cite{9909076}. Due to the degeneracy on the parameter space, cosmic microwave background (CMB) alone can not put the strong constraint on the present value of the Hubble parameter, $H_0$ \cite{0407158}. Thus, it is important to obtain accurate and indepentent bounds on the $H_0$ from many different kinds of observations. 
The Hubble Key Project uses the Hubble Space Telescope (HST) to establish the most precise optical determination of $72 \pm 8$ (km/s/Mpc) \cite{0012376}. The Wide Field Camera 3 (WFC3) on the HST using nearby Cepheids and supernovae (SNe Ia) determines the value $73.8 \pm 2.4$ (km/s/Mpc)  with 3.3 \% uncertainty \cite{11032976}.

The so called ``cosmic chronometers'' approach provides an independent method to constrain $H(z)$ from the differential evolution of massive and passive early type galaxies. One can estimate the expansion rate of the Universe at different epoches using the age difference between two passively evolving old elliptical galaxies \cite{0106145}. By using spectroscopic data of passive early-type galaxies selected from the Sloan Digital Sky Survey (SDSS), the present value of Hubble parameter is obtained as $72.6 \pm 2.9 \pm 2.3$ (km/s/Mpc) \cite{10100831}. The Hubble parameter also can be obtained from the measurements of the angular diameter distance from galaxy clusters via Sunyaev-Zeldovich effect (SZE) combined with measurements of the X-ray flux. In order to obtain the Hubble parameter value from these measurements, one needs to break the degeneracy on the cosmological parameters by applying joint analysis involving the baryonic acoustic oscillations (BAO) or the CMB shift parameter. For a flat $\Lambda$CDM model, the present value is given by $71.4_{-3.4}^{+4.4}$ (km/s/Mpc) \cite{10064200}.

We can also measure the Hubble parameter from the large scale structure (LSS) of the Universe. Both the diameter distance and the Hubble parameter can be measured from the BAO. Using the SDSS III BAO spectroscopic survey from Data Release 9 of 264,283 galaxies, the Hubble parameter value at the effective redshift $z = 0.53$ is measured as $92.9 \pm 7.8$ (km/s/Mpc) \cite{13034666}.

However, there exists tension in the present value of Hubble parameter between the above mentioned methods and the recently released by Planck satellite mission. The Planck CMB only data provides the value $67.4 \pm 1.4$ (km/s/Mpc) \cite{13035076}. Thus, we need to check the consistency of the current measurements of $H$ with various aspects. If we break the assumption that the cosmological constant as a dark energy, then this tension can be released \cite{13047119}. Thus, one can regard the other mechanism to accelerate the current accelerating expansion of the Unvierse like the quintessence models or the modified gravity theories \cite{13046984, 13072002}. The modified gravity theories can produce the different matter growth from the one obtained from $\Lambda$CDM or quintessence models.

The present value of the matter energy density contrast, $\Omo$ is a nuisance cosmological parameter which is prerequisite to measure the Hubble parameter. We will discuss the method to extract the values of both $\Omo$ and the Hubble parameter from the observed redshift space distortions (RSDs) due to the falling of galaxies onto a cluster. In measuring the 2 dimensional power spectrum, the geometric distortions known as the Alcock-Paczynski effect \cite{AP} - distances measured along the line of sight look different to those measured perpendicular to the line of sight are degenerated with the RSDs. The lifting this degeneracy have been considered \cite{11112544, 12031002} and it is a prerequisite for adopting our method to extract the $\Omo$ and the Hubble parameter. We will show how to get both $\Omo$ and the Hubble parameter from the values of the normalized growth rate, $f \sigma_8$ measured from the matter power spectrum or two-point correlation function in the Sec. II. In Sec III, we will estimate the current and future constrain on the growth rate. We show the differences from the different models in Sec IV. We conclude in the last section.

\section{Measuring the matter energy density contrast and the Hubble parameter}
\setcounter{equation}{0}

In this section, we derive the required formulas for both the matter energy density contrast and the Hubble parameter from the measurement of the redshift space distortions (RSDs). We will investigate the errors on both quantities from various aspects. If we consider the Einstein's general relativity (GR) is true, then during the galaxy formation, all modes of overdensity perturbation inside the horizon evolve identically. Thus, the linear perturbation of the matter ($\delta_{\m} = \delta \rho_{\m} / \rho_{\m}$)  on the sub-horizon scales where the dark energy is homogeneous is governed by \cite{Bonnor}
\be \ddot{\delta}_{\m} + 2 H \dot{\delta}_{\m} = 4 \pi G \rho_{\m} \delta_{\m} \, , \label{deltamt} \ee where dot means the derivatives with respect to the cosmic time $t$, $\rho_{\m}$ is the mean matter density, $H$ is the Hubble parameter given by the Friedmann equation ($H^2 = \fr{8 \pi G}{3} \rho_{cr}$), and $G$ is the Newtonian gravitational constant obtained from the Poisson equation. The same technique was used to constrain the time variation of the Newton's constant using WiggleZ measurement \cite{11073659}. However, we will show that the high redshift measurements are required to use this method. This equation is valid at late times and there exists the exact analytic solution for the constant equation of the state of the dark energy \cite{09061643, 09072108}. The growing mode solution of $\delta_{\m}$ is defined as the linear growth factor $g$. The growth factor satisfies the initial conditions $g(a_i) = a_i$ and $\fr{dg}{da} |_{(a=a_i)} = 1$ where $a_i$ is the initial scale factor. If we change the variable in Eq. (\ref{deltamt}) from $t$ to the scale factor, $a$ then we obtain a first order non-homogeneous linear differential equation for $H^2(a)$
\be \fr{d H^2}{d a} + \fr{d \ln [a^6 (\fr{dg}{da})^2]}{da} H^2 = 3 \Omo H_{0}^2 g \Bigl( \fr{d g}{da} \Bigr)^{-1} a^{-5} \, , \label{Ha1} \ee where $H_0$ is the present value of the Hubble parameter and the present value of the matter energy density contrast is defined as $\Omo = \rho_{\m 0} / \rho_{\rm{cr} 0}$. The solution for the above equation (\ref{Ha1}) is given by \cite{9810431}
\be H^2(a) \simeq \fr{3 \Omega_{\m 0} H_0^2}{a^6} \Bigl(\fr{d g}{da} \Bigr)^{-2} \int_{a_{\ast}}^{a} a' g \fr{d g}{da'} da' = \fr{3 \Omega_{\m 0} H_0^2}{a^4} (g f )^{-2} \int_{a_{\ast}}^{a} g^2 f da'  \, , \label{Ha2} \ee where the growth rate is defined as $f \equiv \fr{d \ln g}{d \ln a}$ and $a_{\ast}$ is an arbitrary initial value of the scale factor which satisfies $a_{\ast}  \geq a_{i}$. We emphasize that the initial value $a_{i} \neq 0$ because the sub-horizon equation (\ref{deltamt}) is only valid at late time. We obtain the approximate solution in the above equation (\ref{deltamt}) because we use $a_{\ast}$ in the solution. If we replace $a_{\ast}$ by $a_{i}$, then we recover the equality in the above Eq. (\ref{Ha2}). One can obtain $\Omo$ by inserting $a=1$ in the above equation (\ref{Ha2})
\be \Omo \simeq (g_{0} f_{0})^2 \Bigl( 3 \int_{a_{\ast}}^{1} g^2 f da \Bigr)^{-1}  = (\sigma_8 f_0)^2 \Bigl( 3 \int_{0}^{z_{\ast}} \fr{\sigma_8(z)^2 f(z)}{(1+z)^2} dz \Bigr)^{-1}\, , \label{Omo} \ee where $g_{0}$ and $f_0$ are the present values of $g$ and $f$, respectively. In the second equality, we also use the fact that $\sigma_8(z) = \fr{g(z)}{g_0} \sigma_8$ where $\sigma_8$ is the present value of root-mean-square amplitude of the linear mass density fluctuation on a smoothing sphere of $8 h^{-1}$Mpc radius. We represent the above equation (\ref{Omo}) by using $\sigma_8(z) f(z)$ because they can be measured by the RSDs from the matter power spectrum or the two point correlation function.

\subsection{Errors on the matter energy density contrast}

\begin{center}
\begin{figure}
\vspace{1.5cm}
\centerline{\epsfig{file=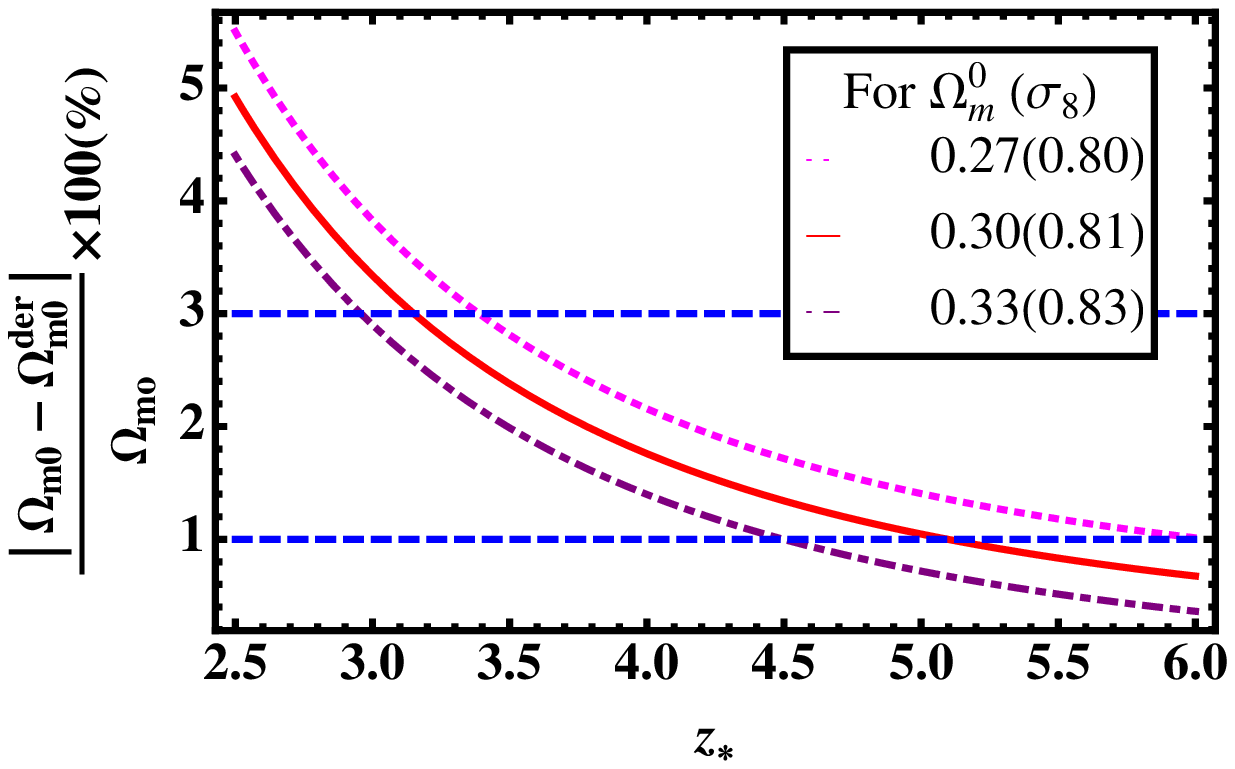, width=7.0cm} \epsfig{file=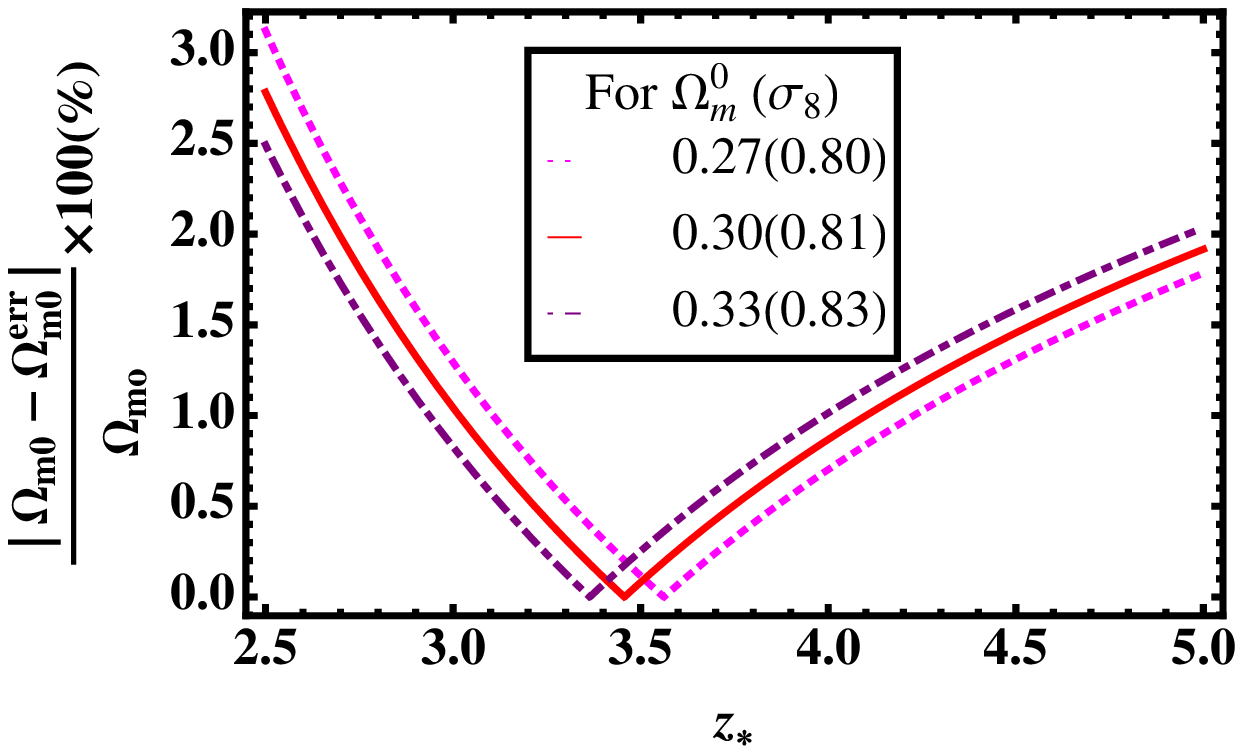, width=7.0cm}}
\vspace{-0.5cm}
\caption{ We show the errors of the derived values of $\Omo$ obtained from Eq. (\ref{Omo}). a) The errors on the derived values of the present matter density contrast, $\Omo^{\rm{der}}$ compared to the true $\Omo$ value for the different values of $z_{\ast}$. $3$ \% errors are obtained at $z_{\ast} = 3.5, 3.2$, and $3.0$ for the fiducial values of $(\Omo, \sigma_8) = (0.27, 0.80), (0.30, 0.81)$, and $(0.33, 0.83)$, respectively.  $1$ \% errors are achieved at $z_{\ast} = 6.0, 5.2$ and $4.5$ for the sames values of $(\Omo, \sigma_8)$, respectively. b) The differences between the value of $\Omo^{\rm{der}}$ with $\sqrt{1+z}$ \% errors on $f \sigma_8$ and $\sigma_8$, $\Omo^{\rm{err}}$ and the true $\Omo$ value for the different values of $z_{\ast}$.} \label{fig1}
\end{figure}
\end{center}

When $\Omo$ is measured by using the Eq. (\ref{Omo}), we need to distinguish two main sources of its errors. First one is an intrinsic error due to the approximate sub-horizon scale solution. The other is due to the measurement errors on $f \sigma_8$ and $\sigma_8$ from the observation. One can understand the first intrinsic error from the fact that the derived value of the present matter density contrast, $\Omo^{\rm{der}}$ approaches to its fiducial value, $\Omo$ when $a_{\ast}$ approaches to $a_{i}$. Thus, this error on $\Omo^{\rm{der}}$ depends on $z_{\ast}$. We show this in the left panel of figure \ref{fig1}. In this section, we assume that the fiducial model is the flat $\Lambda$CDM universe. We show the differences between $\Omo$ and $\Omo^{\rm{der}}$ as a function of $z_{\ast}$ in the left panel of Fig. \ref{fig1}. We show these errors for the different sets of fiducial values of $\Omo$ and $\sigma_8$. The dotted, solid, and dotdashed lines correspond to $(\Omo, \sigma_8) = (0.27, 0.80), (0.30, 0.81)$, and $(0.33, 0.83)$, respectively. In this analysis we adopt $z_{i} = 19$. As expected, the errors are decreased as $z_{\ast}$ increases. The larger the fiducial $\Omo$ value, the smaller the error for the same value of $z_{\ast}$. $3 (1)$ \% errors are obtained at $z_{\ast} = 3.5 (6.0), 3.2 (5.2)$, and $3.0 (4.5)$ for the fiducial values of $(\Omo, \sigma_8) = (0.27, 0.80), (0.30, 0.81)$, and $(0.33, 0.83)$, respectively. We can obtain less than $3$ \% errors when $z_{\ast} > 3.5$ for all the models. We also check the other error on $\Omo^{\rm{der}}$ due to the measuring errors in $f (z) \sigma_8(z)$ and $\sigma_8(z)$. If we assume that the errors on the measurements of both $f \sigma_8$ and $\sigma_8$ as $\sqrt{1+z}$ \%, then the errors on $\Omo^{\rm{der}}$ compared to the fiducial values are less than $1$ \%  at $3.2 \leq z_{\ast} \leq 4.0$ for all the models. This is shown in the right panel of Fig. \ref{fig1}. This figure shows the differences between $\Omo^{\rm{der}}$ with the measurement errors on $f \sigma_8$ and $\sigma_8$ and the fiducial value of $\Omo$.  Again, the dotted, solid, and dotdashed lines correspond to $(\Omo, \sigma_8) = (0.27, 0.80), (0.30, 0.81)$, and $(0.33, 0.83)$, respectively. The induced errors on $\Omo$ can be less than $1$ \% when $z_{\ast}$ are inside $(3.1, 4.2), (3.0, 4.1), (2.95, 4.0)$ for $(\Omo, \sigma_8) = (0.27, 0.80), (0.30, 0.81)$, and $(0.33, 0.83)$, respectively. These facts indicate that the main error on $\Omo^{\rm{der}}$ comes from the approximate solution in Eq. (\ref{Omo}). Thus, this method is useful when one can measure the $f \sigma_8$ up to $z \sim 3$.

\subsection{Errors on the Hubble parameter}

\begin{center}
\begin{figure}
\vspace{1.5cm}
\centerline{\epsfig{file=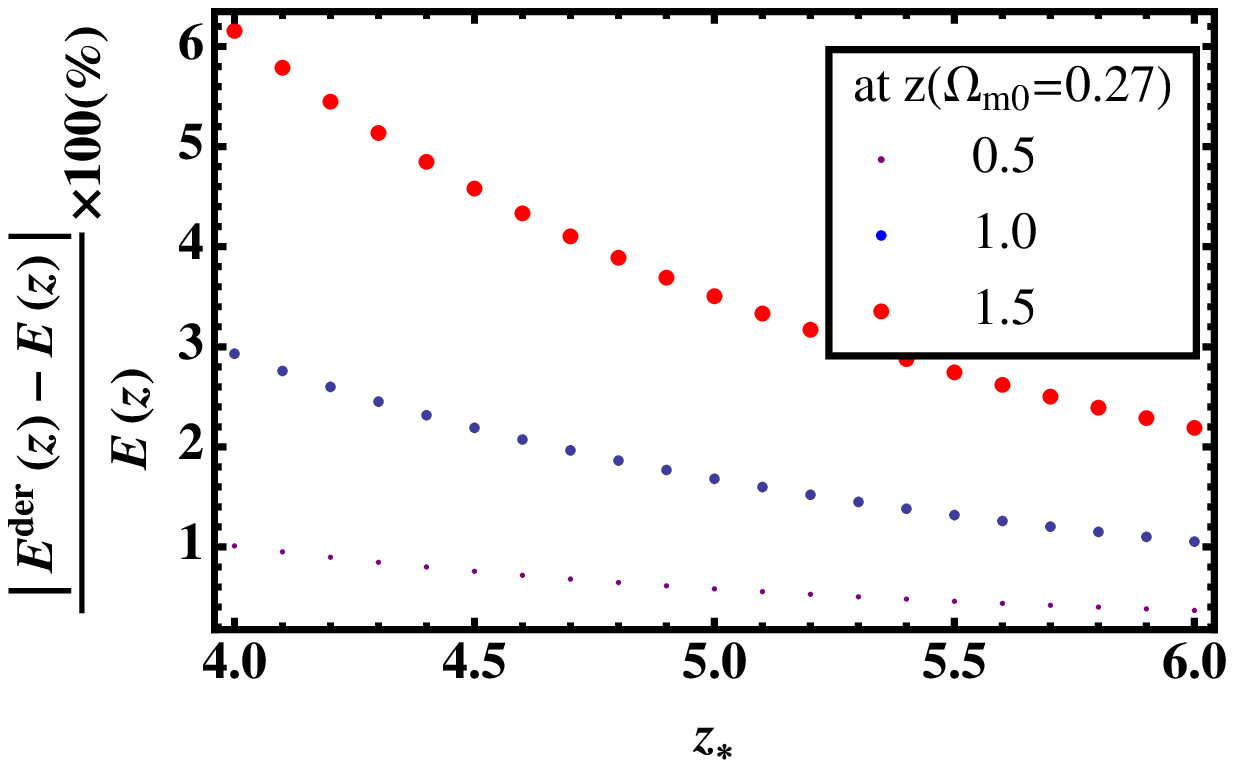, width=7.0cm} \epsfig{file=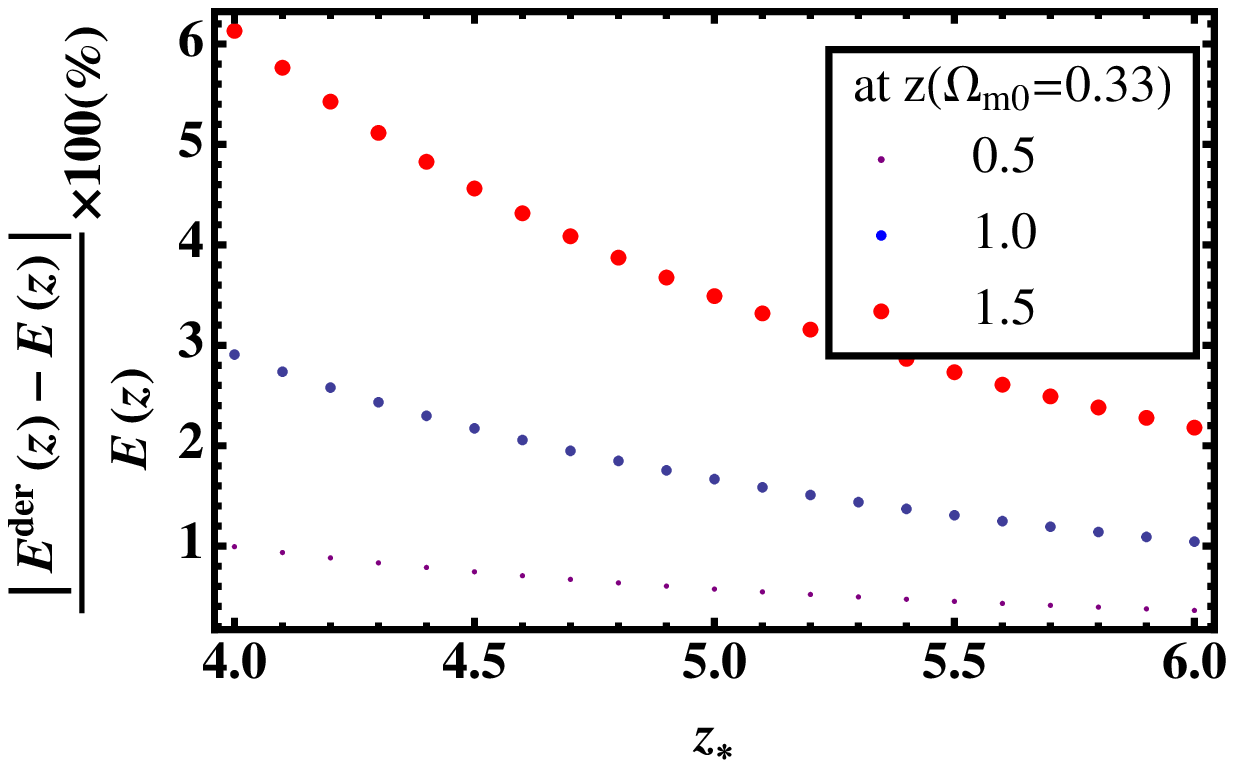, width=7.0cm}}
\vspace{-0.5cm}
\caption{ We show the errors of $E(z)$ for the different values of $\Omo$ as a function of $z_{\ast}$. a) The errors on $E(z)$ with $\Omo = 0.27$ and $\sigma_8 = 0.8$ at different epoch. The small, medium, and large dots correspond to  $z = 0.5, 1.0$, and $1.5$, respectively. $3$ \% errors are obtained at $z_{\ast} = 4$ and $5.5$ for $z = 1.0$ and $1.5$, respectively.  $1$ \% errors are achieved at $z_{\ast} \geq 4$ for $z = 0.5$.  b) The errors on $E(z)$ when $\Omo = 0.33$ and $\sigma = 0.83$ at different epoch. The results are almost indentical to $\Omo = 0.27$ model. } \label{fig2}
\end{figure}
\end{center}

In this subsection, we consider how to obtain the Hubble parameter and also investigate the errors on the Hubble parameter. Now from Eqs. (\ref{Ha2}) and (\ref{Omo}), one can obtain \ba E(a) &\equiv& \fr{H(a)}{H_{0}} \simeq a^{-2} \fr{g_{0} f_0}{g(a) f(a)} \sqrt{\fr{\int_{a_{\ast}}^{a} g^2 f da'}{\int_{a_{\ast}}^{1} g^2 f da'}} = a^{-2}  \fr{\sigma_{8} f_0}{\sigma_{8}(a) f(a)} \sqrt{\fr{\int_{a_{\ast}}^{a} \sigma_{8}(a')^2 f(a') da'}{\int_{a_{\ast}}^{1} \sigma_{8}(a')^2 f(a') da'}} \nonumber \\ &=& a^{-2} \fr{\sigma_{8} f_0}{\sigma_{8}(a) f(a)} \sqrt{1 - \fr{\int_{a}^{1} \sigma_{8}(a')^2 f(a')  da'}{\int_{a_{\ast}}^{1} \sigma_{8}(a')^2 f(a') da'}} \, . \label{HaHo} \ea  It is convenient to rewrite Eq. (\ref{HaHo}) using the redshift $z$,
\be E(z) \equiv \fr{H(z)}{H_0} \simeq (1+z)^{2} \fr{\sigma_{8} f_0}{\sigma_{8}(z) f(z)}  \sqrt{ 1 - \fr{\int_{0}^{z} \fr{\sigma_{8}(z')^2 f(z')}{(1+z')^2}  dz'}{\int_{0}^{z_{\ast}} \fr{\sigma_{8}(z')^2 f(z')}{(1+z')^2} dz'} } \, . \label{HzHo} \ee Thus, one can estimate $E(z)$ by using the model independent growth factor $f(z) \sigma_8(z)$ measurements. There are several important aspects in the above equation (\ref{HzHo}). First, Eq. (\ref{HzHo}) is true for any dark energy model. It includes $\Lambda$CDM, quintessence models, and phantom models. It means that this equation is independent of the dark energy equation of state, $\omega$. If one considers the so called ``Chevallier-Polarsky-Linder (CPL)'' parametrization \cite{0009008, 0208512}, then $E(z)$ is represented by
\be E(z) = \sqrt{\Omo (1+z)^3 + (1 - \Omo) (1+z)^{3(1+\omega_0+\omega_a)} \exp \Bigl[-3 \omega_a \fr{z}{1+z} \Bigr]} \, , \label{ECPL} \ee where $\omega = \omega_0 + \omega_a \fr{z}{1+z}$. Because our fiducial model is a flat $\Lambda$CDM, we can put $\omega_0 = -1$ and $\omega_a = 0$. Second, this equation is also independent of the current value of the Hubble parameter, $H_0$. Thus, one does not need to rely on the other observations for this value. Also the error on $E(z)$ is almost independent of the fiducial value of $\Omo$. This will be shown below. Last interesting point is that one can measure the $E(z)$ values at the different epochs. Thus, this method is quite useful in order to compare with the data from the standard ruler method like BAO which can measure $H(z)$ at the specific epoch. However, there is a limitation of the accuracy on $E(z)$ because of $z_{\ast}$ dependence. There also exist two errors on $E(z)$ derived from Eq. (\ref{HzHo}). We analyze the errors on $E(z)$ at the different epoch as a function of $z_{\ast}$ and $\Omo$. These are shown in Fig.\ref{fig2}. Again our fiducial model is a flat $\Lambda$CDM and we can compute the error on $E(z)$ obtained from Eq. (\ref{HzHo}) compared to the fiducial value from Eq. (\ref{ECPL}). In the left panel of Fig.\ref{fig2}, we show the errors on $E(z)$ at the different epoch as a function of $z_{\ast}$ when we adopt $\Omo = 0.27$ and $\sigma_8 = 0.8$. The small, medium, and large dots (from bottom to top) correspond to the errors on $E^{\rm{der}}(z)$ at $z = 0.5, 1.0$, and $1.5$, respectively. For $z \leq 1.0$, errors are smaller than $3$ \% as long as $z_{\ast} \geq 4$. In order to measure $E(z)$ with $3$ \% level at $z =1.5$, one needs to extend $z_{\ast} \simeq 5.5$. We also show the same plot for $\Omo = 0.33$ and $\sigma_8 = 0.83$ in the right panel of Fig.\ref{fig2}. It shows the almost same behavior as the $\Omo = 0.27$ model. We also consider the effect of the measurement errors of $f(z) \sigma_8(z)$ on $E(z)$. It provides less than percent errors on $E(z)$ when we consider $\sqrt{1+z}$ \% measurement errors on $f(z) \sigma_8(z)$. This is due to the fact that the effect from the errors on $f(z) \sigma_8(z)$ both inside and outside of the square roots are compensated from both numerator and denominator of Eq. (\ref{HzHo}). Thus, one can safely ignore the effect of measurement error. However, this is a systematic error related to our method and it can be added in quadrature to the statistical measurement error.

\subsection{Estimation on the Hubble parameter from observation}

\begin{center}
\begin{figure}
\vspace{1.5cm}
\centerline{\epsfig{file=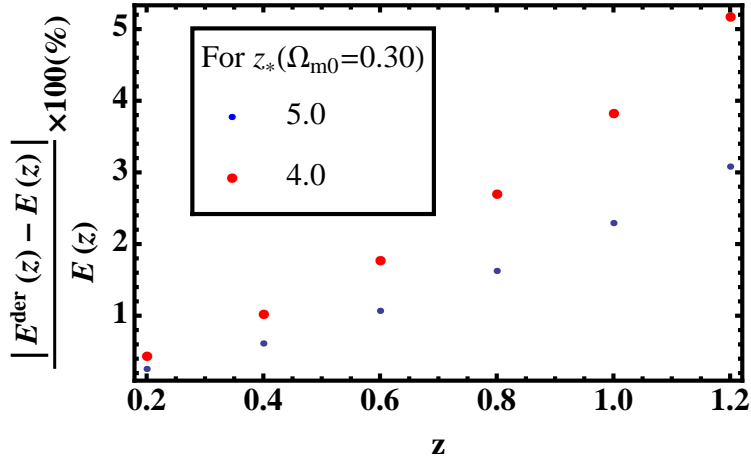, width=10.0cm} }
\vspace{-0.5cm}
\caption{ We show the errors on the derived value of $E(z)$ by using the trapezoidal rule with $\Delta z = 0.2$
for the different values of $z_{\ast}$ at the different epoch. When $z_{\ast} = 5$, all the values of $E^{\rm der}(z)$ have than less than $3$ \% error for $z \leq 1.2$. The errors on $E^{\rm der}(z)$ at $z = 1.2$ can be as large as $5$ \% when $z_{\ast} = 4$. } \label{fig3}
\end{figure}
\end{center}

Based on the real observation, the integral terms in the above equations should be replaced by the so called ``trapezoidal rule''
\ba \int_{a}^{b} f(x) dx &\simeq& \fr{1}{2} \sum_{j=1}^{N} (x_{j+1} - x_{j}) \Bigl[ f(x_{j+1}) + f(x_j) \Bigr] \,\, , {\rm in\, non-uniform\, grid} \label{trapezoidalnonunif} \\
&\simeq& \fr{h}{2} \sum_{j=1}^{N} \Bigl[ f(x_{j+1}) + f(x_j) \Bigr] = \fr{b-a}{2N} \Bigl[ f(a) + 2 f(x_2) + \cdots + 2 f(x_{N-1}) + f(b) \Bigr] \, , \nonumber \\ && {\rm in\, uniform\, grid} \label{trapezoidalunif}  \ea
One needs to compare the error on $E(z)$ induced by changing the integral in Eq. (\ref{HzHo}) into the sum in the above equations (\ref{trapezoidalnonunif}) or (\ref{trapezoidalunif}). Again the error on $E(z)$ depend on $z_{\ast}$ and this is shown in Fig. \ref{fig3}. In this figure, we use the uniform grid with redshift bin $\Delta z = 0.2$. When we use $z_{\ast} = 5.0$ with $(\Omo, \sigma_8) = (0.30, 0.81)$, the error on $E(z)$ compared to the fiducial value is about $2.2$ \% at $z = 1.0$. When we use the integration, the error on $E(z)$ at $z = 1.0$ is about $1.8$ \%. If we use $z_{\ast} = 4.0$ for the same set of $(\Omo, \sigma_8)$, then error on $E(z)$ is about $3.8$ \% which has about $3$ \% error when we use the integration. Even if we use the non-uniform grid, the induced error on $E(z)$ is not changed much compared to uniform case. When we choose the bin size as $\Delta z = 0.1$, then the errors on $E(z)$ at $z = 1.0$ are about $2$ and $3.3$ \% for $z_{\ast} = 5$, and $4$. Thus, if one uses the bin interval as $\Delta z = 0.1$, then the error induced by the summation of the measurement will be ignorable.

We show that the measurement errors on $f \sigma_8$ have negligible effect on estimating the values of  $E^{\rm{der}}(z)$. Thus, one can compare the measurement of $H(z)$ obtained from the perturbation quantities using Eq. (\ref{HzHo}) against the one from the background evolution by Eq. (\ref{ECPL}). As mentioned in the previous subsection, this is a systematic error related to our method and it can be added in quadrature to the statistical measurement error.

\section{Constraints on Growth Rate}
\setcounter{equation}{0}

The effect of linear RSDs on the two point correlation function is described by \cite{Kaiser}
\be \xi_{s}(\sigma, \pi) = \Bigl(1 + \beta \fr{\partial^2}{\partial \pi^2} (\nabla^2)^{-1}\Bigr)^2 \xi_{r}(r) \, , \label{Ps} \ee where $\xi_s$ and $\xi_r$ are the redshift space and real space galaxy two point correlation functions at distance $r$ and $\beta = \fr{f}{b}$ represents the amplitude of RSDs in terms of the growth rate $f$ and the linear galaxy bias factor $b$. The growth rate $f$ is defined as $f = \fr{d \ln g}{d \ln a}$. However, the measured value of $b$ is degenerate with $\sigma_8$. Thus, $f \sigma_8$ can be used as a model independent value. There have been several methods to measure the growth rate in a galaxy redshift survey. Because the peculiar velocities are directly proportional to the growth rate $f$, one can derive the value of $f$ by measuring the peculiar velocities \cite{12034814}. One can also use the QSO clustering and Ly-$\alpha$ clustering to measure the correlation function. From the correlation function, the growth rate can be derived by using the observed value of $\beta$ and $b$ \cite{0612401}. Also the growth rate can be obtained from the RSDs measurements.

\subsection{Current constraints}

The most current measurements of the growth rate are obtained by using RSDs. Using LasDamas simulations of the SDSS-II Luminous Red Galaxy (LRG) data, clustering in these mock catalogues is recovered on scales $30 \sim 200$ $\rm{h}^{-1}$ Mpc \cite{11021014}. From this, the model independent measurements are given by $f(z=0.25)\sigma_8(z=0.25) = 0.3512 \pm 0.0583$ and $f(z=0.37)\sigma_8(z=0.37) = 0.4602 \pm 0.0378$. SDSS-III Baryon Oscillation Spectroscopic Survey (BOSS) using a passively evolving population of galaxies also provides the growth rate at four different epoches \cite{12036565}. Measurements are given by $f(z=0.3)\sigma_8(z=0.3) = 0.407 \pm 0.055$, $f(z=0.4)\sigma_8(z=0.4) = 0.419 \pm 0.041$, $f(z=0.5)\sigma_8(z=0.5) = 0.427 \pm 0.043$, and $f(z=0.6)\sigma_8(z=0.6) = 0.433 \pm 0.067$. The analysis of RSD in the two-point correlation function of the 6dF Galaxy Survey (GS) measure the value at $z = 0.067$ as $f(z=0.067)\sigma_8(z=0.067) = 0.423 \pm 0.055$ \cite{12044725}. From the WiggleZ Dark Energy Survey two-point correlation function analysis by using the Quadratic Correlation Function model, $f \sigma_8$ values at four different epoches are obtained as $f(z=0.2)\sigma_8(z=0.2) = 0.40 \pm 0.13$, $f(z=0.4)\sigma_8(z=0.4) = 0.39 \pm 0.08$, $f(z=0.6)\sigma_8(z=0.6) = 0.40 \pm 0.07$, and $f(z=0.76)\sigma_8(z=0.76) = 0.46 \pm 0.10$ \cite{13025178}. The VIMOS Public Extragalactic Redshift Survey (VIPERS) provides $f(z=0.8) \sigma_8(z=0.8) = 0.47 \pm 0.08$ \cite{13032622}. From Data Release 9 (DR9) CMASS sample of the SDSS-III BOSS, $f(z=0.57) \sigma_8(z=0.57) = 0.428 \pm 0.066$ is given \cite{13034486}. We summarize these in Table \ref{table1}. The errors on $f \sigma_8$ are about $10 \sim 20$ \% and this will not be a matter in our analysis to obtain $E(z)$. However, the redshift range for the current measurement is less than $1$ and this will produce large error in $E(z)$ by using our method. One needs at least $z > 3$ galaxy survey to apply the method.

\begin{center}
    \begin{table}
    \begin{tabular}{ | c | c | c | c | c | c | c | c |  }
    \hline
	Survey & z & $f \sigma_8$ & Ref & Survey & z & $f \sigma_8$ & Ref  \\ \hline
	SDSS-II & $0.25$ & $0.3512 \pm 0.0583$ & \cite{11021014} &  & $0.2$ & $0.40 \pm 0.13$ & \\ \cline{2-3}\cline{6-7}
	LRG	    & $0.37$ & $0.4602 \pm 0.0378$ &  & WiggleZ & $0.4$ & $0.39 \pm 0.08$ & \cite{13025178}  \\ \cline{1-4}\cline{6-7}
 	SDSS-III & $0.3$ & $0.407 \pm 0.055$ &  &   & $0.6$ & $0.40 \pm 0.07$ & \\ \cline{2-3}\cline{6-7}
	BOSS	     & $0.4$ & $0.419 \pm 0.041$ &  \cite{12036565} &  & $0.76$ & $0.46 \pm 0.10$ & \\ \cline{2-3}\cline{5-8}
		     & $0.5$ & $0.427 \pm 0.043$ & & VIPERS & $0.8$ & $0.47 \pm 0.08$ &  \cite{13032622} \\ \cline{2-3}\cline{5-8}
	             & $0.6$ & $0.433 \pm 0.067$ & & CMASS & $0.57$ & $0.428 \pm 0.066$ &  \cite{13034486} \\ \hline
	6dF GS & 0.067 & $0.423 \pm 0.055$ & \cite{12044725}  & \multicolumn{4}{l|}{} \\ \hline
    \end{tabular}
    \caption{Summary for the observational data of $f \sigma_8$ in the current galaxy surveys. }
    \label{table1}
    \end{table}
\end{center}


\subsection{Future constraints}
We calculate expected errors of $E(z)$ measurement for future experiment. We need survey at least $z \simeq 3.0$ in order to obtain the proper value of $\Omo$ as shown before. Even though some surveys will not reach to this redshift, we can still use the above equation (\ref{Omo}) as long as the fiducial model is $\Lambda$CDM. This is due to the fact that the growth function will converge to $g(z) = (1+z)^{-1}$ from $z > 2$. The relative error on $f \sigma_8$ can be expressed as  \cite{08021944}
\be \fr{\Delta (f \sigma_8)}{f \sigma_8} = \fr{50}{(r \langle n_g \rangle )^{0.44} \sqrt{V}} \label{errorfsigma8} \, , \ee
where $r  \simeq 0.2$ is the sampling rate, $\langle n_g \rangle $ is the galaxy number density in $(h \rm{Mpc}^{-1})^3$, and $V$ is the survey volume in  $(h^{-1} \rm{Mpc})^3$. Wide-Field Fiber-Fed Optical Multi-Object Spectrograph (WFMOS) for the Gemini/Subaru observatory is proposed to measure the acoustic oscillations at $0.5 < z < 1.3$ and $2.5 < z < 3.5$ using a redshift survey of millions of galaxies over 2000 square degrees. Big Baryon Oscillation Spectroscopic Survey (BigBOSS) is a ground based dark energy experiment to study BAO and the growth of structure with an all sky galaxy redshift survey.
Euclid is a planned space telescope scheduled to be launched in 2020 with 6 years mission length. Its goal is to map the large scale distribution of dark matter and characterize properties of dark energy. It is the result of the merging between Dark Universe Explorer (DUNE, intended to measure effects of weak gravitational lensing) and Spectroscopic All Sky Cosmic Explorer (SPACE, to measure the baryon acoustic oscillations). The Large Synoptic Survey Telescope (LSST) is a planned wide-field survey reflecting telescope that will photograph the entire available sky every few nights. The LSST is currently in its design and development phase and full operations for a ten-year survey commencing January 2022. The survey redshift range, number of galaxies, survey volume, and expected error on $f \sigma_8$ are shown in Table \ref{table2}. Both WFMOS and LSST are suitable for obtaining $E(z)$ by using the mentioned method.

\begin{center}
    \begin{table}
    \begin{tabular}{ | c | c | c | c | c | }
    \hline
	Survey & z & N & V $(h^{-1} {\rm Mpc})^{3}$ & $\Delta (f \sigma_8) / (f \sigma_8)$ (\%) \\ \hline
	WFMOS & $0.5 \sim 1.3$ & $2 \times 10^{6}$ & $4 \times 10^{9}$ & 4.5 \\ \cline{2-5}
		     & $2.3 \sim 3.3$ & $6 \times 10^{5}$ & $1.2 \times 10^{9}$ & 8.3 \\ \hline
 	BigBOSS & $0.2 \sim 2$ & $3 \times 10^{7}$ & $3 \times 10^{10}$ & 1.2 \\ \hline
	EUCLID & 2 & $5 \times 10^{8}$ & $1 \times 10^{11}$ & 0.3 \\ \hline
	LSST & $0.15 \sim 3$ & $1 \times 10^{10}$ & $1 \times 10^{11}$ & 0.1 \\ \hline
    \end{tabular}
    \caption{Summary for the on-going and future galaxy surveys. $z$ is the covering redshift, $N$ is the measuring galaxy number, $V$ is the survey volume in $(h^{-1} {\rm Mpc})^{3}$ unit, and $\Delta (f \sigma_8) / (f \sigma_8)$ is the estimated relative error on the measurement $f \sigma_8$. }
    \label{table2}
    \end{table}
\end{center}


\section{Model Comparison}
\setcounter{equation}{0}

Recently, the alternative gravity theories are investigated as a origin of the accelerating expansion universe.
One can compare the values of $f(z) \sigma_8(z)$ for the different viable models. We consider the $\Lambda$CDM, general $f(R)$,  scalar tensor gravity (STG), and Dvali-Gabadadze-Porrati (DGP) models \cite{0005016}. If one considers the modified gravity theories, then the theoretical prediction of the evolution of the growth factor $g(a)$ on sub-horizon scales given in Eq. (\ref{deltamt}) is modified as
\be g''(k,a) + \Bigl( \fr{3}{a} + \fr{H'(a)}{H(a)} \Bigr) g'(k,a) - \fr{3}{2} \fr{\Omo}{a^2 E^2(a)} p(k,a) g(k,a) = 0 \label{modg} \ee where primes denote the derivatives with respect to the scale factor $a$ and $p(k,a)$ depends on the particular form of the dynamical equations of the gravity theory considered. In $\Lambda$CDM model, $p(k,a) = 1$. The alternative models that we consider in this work are : \\

\hspace{0.1 in} $\bullet$ f(R) model. The general $f(R)$ modified gravity theory can be described by the following action \be S = \int d^4 x \sqrt{-g} \Bigl( \fr{1}{16 \pi G_{\ast}} f(R) + {\cal L}_m \Bigr) \, \label{SfR} \ee where $G_{\ast}$ is a bare gravitational constant and $f(R)$ is a general function of the Ricci scalar $R$. However, many $f(R)$ models cannot account for a viable cosmic expansion history and a late time accelerating expansion simultaneously \cite{0612180}. However, one might be able to obtain the viable models if one adopts the general form of $f(R)$ models \cite{0610213,10110544}. One of the viable $f(R)$ models suggested by Starobinsky provides $p(k,a) = 1 + \fr{\fr{k^2}{a^2 H^2}}{1 + 3\fr{k^2}{a^2 H^2}}$ and can be approximated as $p(k,a) \simeq \fr{4}{3}$ at sub-horizon scales \cite{07062041}.  \\

\hspace{0.1 in} $\bullet$ STG model. STG are described in the Jordan frame by \be S = \fr{1}{16 \pi G_{\ast}} \int d^4 x \sqrt{-g} \Bigl( F(\phi) R -\nabla^{\mu} \phi \nabla_{\mu} \phi - 2 U(\phi) \Bigr) + S_{\rm{m}} (g_{\mu\nu} , \psi_{\rm{m}} \Bigr) \, \label{SSTG} \ee where $F(\phi)$ is a dimensionless function and $S_{\rm{m}}$ is a matter action for matter fields. In STG models, one obtains \be p(k,a) = \fr{G_{\rm{eff}}(a)}{G_{\rm{eff}}(a=1)} \Bigl( 1 + \fr{1}{1 + \fr{k}{m a}} \Bigr) \simeq \fr{F(a=1)}{F(a)} \, \label{pSTG} \ee where $m$ is the mass of the scalar field $\phi$ \cite{10122646}. We use a simple ansatz for the effective Newton's constant $G_{\rm{eff}}(z) = 1 + \xi \Bigl(\fr{z}{1+z} \Bigr)^2$ with $\xi = -0.2$ in the analysis. \\

\hspace{0.1 in} $\bullet$ DGP model. In DGP model, the Hubble parameter is described by the size of the extra dimension $r_{c}$ \be E(z) = \sqrt{\Omega_{r_{c}}} + \sqrt{\Omega_{r_{c}} + \Omo (1+z)^3} \, \label{EDGP} \ee where $\Omega_{r_c} = \fr{1}{4} ( 1 - \Omo)^2$. And $p(k,a)$ is given by \be p(k,a) = 1 + \fr{1}{3 \beta} \, \label{pDGP} \ee where $\beta = 1 - \fr{E(a)}{\sqrt{\Omega_{r_c}}} \Bigl( 1 + \fr{a}{3} \fr{H'(a)}{H(a)} \Bigr)$.
In the analysis, we choose the same background evolution $H(a)$ for $\Lambda$CDM, f(R), and STG. DGP shows the deviation of the evolution of $H(a)$ as large as $10$ \% from $H(a)$ when we choose the same value of $\Omo$ as other models.

\begin{center}
\begin{figure}
\vspace{1.5cm}
\centerline{\epsfig{file=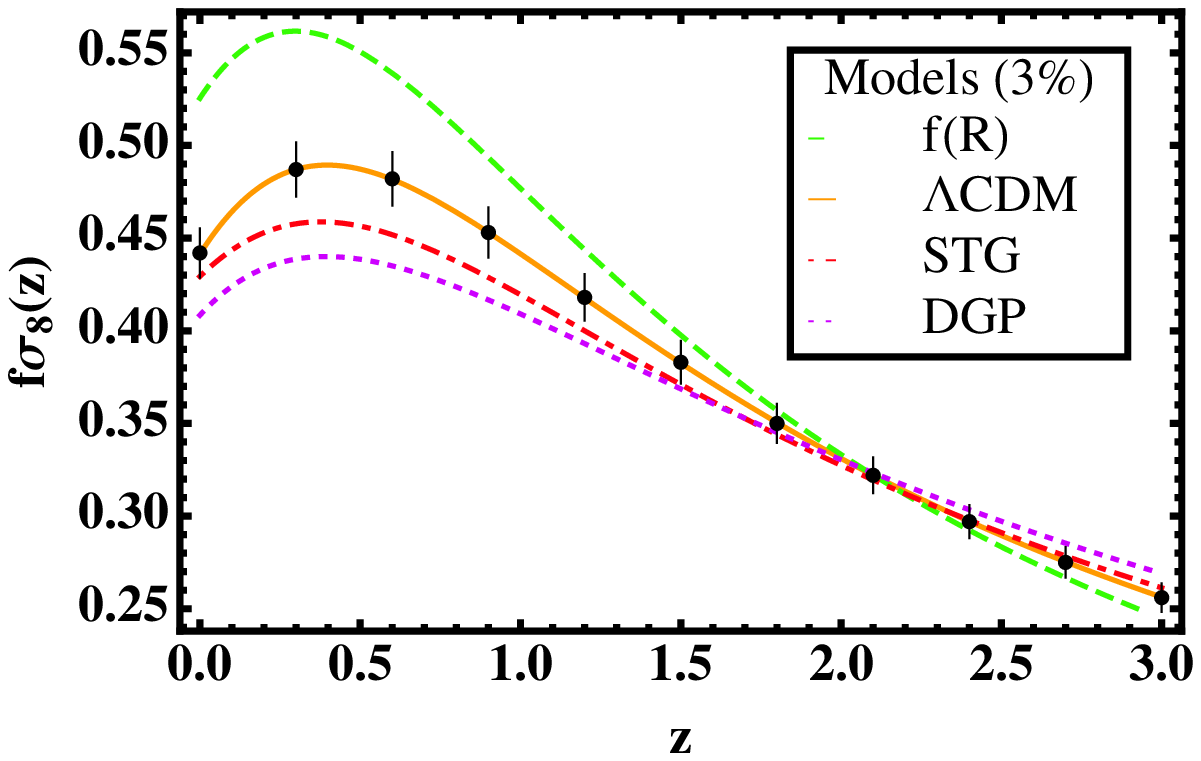, width=7.0cm} \epsfig{file=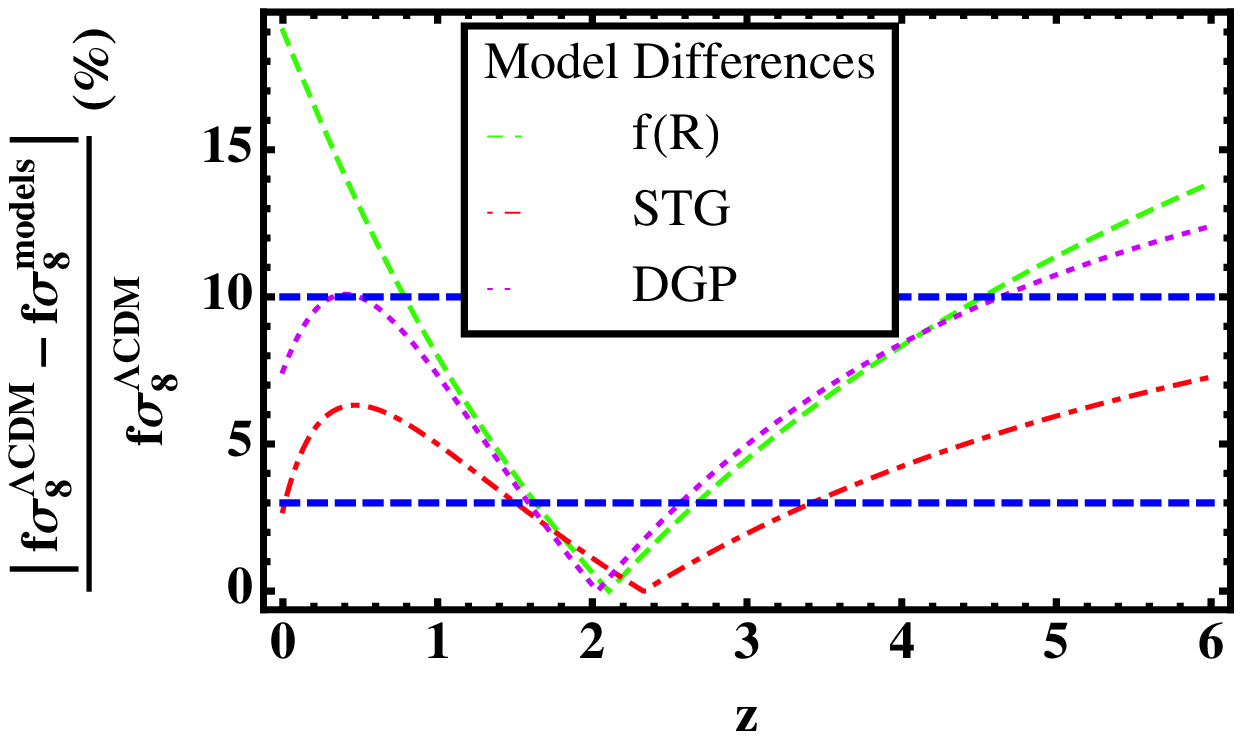, width=7.0cm}}
\vspace{-0.5cm}
\caption{ a) We show the values of $f(z) \sigma_8(z)$ for the different models in the left panel. The $3$ \% error for the fiducial $\Lambda$CDM model is also plotted. The dashed, solid, dotdashed, and dotted lines correspond to f(R), $\Lambda$CDM, STG, and DGP model, respectively. The background evolution, $H(z)$ are same for all models except DGP. b) The differences of $f(z) \sigma_8(z)$ between different models and $\Lambda$CDM model are shown as a function of $z$. The dashed, dotdashed, and dotted lines correspond to f(R), STG, and DGP, respectively. With the $10$ \% measurement errors, we hardly can distinguish the different models with RSDs. } \label{fig4}
\end{figure}
\end{center}

Data is produced for a fiducial $\Lambda$CDM model with $\Omega_{m}^{0} = 0.32$ and $\sigma_8 = 0.83$ to be consistent with Planck data and errors on $f(z) \sigma_8(z)$ are $3$ \%. This is shown in the left panel of Fig. \ref{fig4}. The dashed, solid, dotdashed, and dotted lines correspond to f(R), $\Lambda$CDM, STG, and DGP models, respectively. Even though the background evolution, $H(z)$ are same for all models except DGP, each model produces the different behavior of $f \sigma_8$. This is due to the fact that there exists the modification for the evolution equation of the growth factor as shown in Eq. (\ref{modg}). If the errors on $f(z) \sigma_8(z)$ are about $10$ \% then we will not be able to distinguish the different models based on this observation. This is shown in the right panel of Fig. \ref{fig4}. We show the deviation of $f \sigma_8$ values of $\Lambda$CDM from f(R) (dashed), STG (dotdashed), and DGP (dotted) models in this figure. In future survey, the errors on the observable can be less than percent level as shown in the previous section. Now we can distinguish the different models for $z \leq 1.5$. In higher $z$, the differences of $f \sigma_8$ for the different models become weaker.

\section*{Conclusion}
We investigate the method to measure both the matter energy density contrast and the Hubble parameter directly from the measurements of the normalized growth rate $f \sigma_8$. This approach based on the evolution of the matter perturbation on the sub-horizon scales. Thus, one needs to extend the measurement of the growth rate at early epoch in order to obtain the more accurate solution for the growth rate. However, there exists the limitation to measure the galaxy in the early epoch and the accuracy of this method depends on the maximum redshift of the measurement. We found that the error on the measurement of the energy contrast of the matter becomes less than $3$ \% as long as the growth rate measurement can be reach to $z_{\ast} \geq 3.5$ for all of viable values of $\Omo$ ($\Omo \geq 0.27$). From this method, the nuisance cosmological parameter $\Omo$ can be obtained without assuming any prior.

We can measure the Hubble parameter for $0 < z \leq 1$ with less than $3$ \% when the growth rate measurement can be reach to $z_{\ast} \geq 4$.  For the same measurement, Hubble parameter can be measured less than $6$ \% error for $z \leq 1.5$. The error caused by the measuring errors on the growth rate is negligible and this method is very useful.

For the same background evolution of the Universe, the growth rate can be changed depending on the gravity theories. This method can be used as a complementary in addition to the growth rate to reveal the origin of the current accelerating expansion of the Universe.

\section*{Acknowledgments}
We thanks Xiao-Dong Li and Changbom Park for stimulating discussion. We specially thanks to the anonymous referee for the useful comments.
We also thank KIAS Center for Advanced Computation for providing computing resources.

\end{document}